\documentclass[a4paper,11pt]{article}
\usepackage{amsmath, graphics}

\usepackage{pslatex}
\usepackage{graphicx}
\usepackage[usenames]{color}
\begin{document}

\title{ Ground state and excitation spectra of a strongly correlated lattice  by the coupled cluster method.}

\author{Alessandro Mirone  \\
  European Synchrotron Radiation Facility, BP 220, F-38043 Grenoble Cedex, France}
\maketitle

\abstract

We apply Coupled Cluster Method to a strongly correlated lattice Hamiltonian
and we extend the Coupled Cluster linear response method to the calculation
of electronic spectra.   We do so by    finding an approximation to a resolvent operator which describes 
the spectral response of  the  Coupled Cluster solution 
to  excitation operators.
In our Spectral Coupled Cluster Method the ground and excited states appear
as resonances in the spectra  and the resolvent can be iteratively improved   
in selected spectral regions.
We apply our method to a $MnO_2$ plane model  which 
corresponds to previous experimental works.



\section{Introduction}

  The numerical methods for solid state physics span a wide range of techniques 
which aim to provide approximate solutions
 to the problem of many body interactions in correlated  systems; the exact solution  being unknown. 
Between these techniques some of the
most notables are: the DMF theory, which uses  a self-energy correction term obtained 
from an Anderson impurity many-body solution,   the GW equation, which calculates the self-energy with neglection of vertex corrections,
the quantum monte-carlo method and the Coupled-Cluster Method (CCM). 
The Coupled-Cluster method has been conceived   in the $50s$ by Fritz Coester and Hermann Kummel for nuclear physics, and has 
since progressively gained other domains\cite{reviews}.  In quantum-chemistry, in particular, CCM is widely regarded as the most reliable choice
when high accuracy is needed\cite{ccmchemistry}. 
Concerning the  lattice models of strongly interacting electrons CCM has recently been applied to spin lattices\cite{bishop} and to Hubbard model\cite{roger}. 

 Although the CCM was  initially formulated as a  ground state approximation, the recent developpment of CCM time dependent linear response \cite{linearresp} has stretched the CCM applicability to excited states. In particular  Crawford  and Ruud  have calculated vibrational eigenstates  contributions to Raman optical activity\cite{ramspectra} while  Govind et al.  have calculated excitonic states in potassium bromide\cite{excitons}.

In this paper  we extend the coupled cluster linear response method to the calculation of electronic excitation spectra.
To do so we   represent an  initial wave-function as the product of the probe operator times the CCM  solution, and we develop an  original solution method  of the resolvent equation for the CCM ansatz. 

This paper is organised in the following way. In section \ref{metodo} we detail the equations. In section \ref{modello} we describe the model of a $MnO_2$ plane  derived from previous  absorption and scattering x-ray spectroscopy studies, on which we  test our method. We discuss the results of our Spectral Coupled Cluster Method  in section \ref{discussione}.
There we  also validate the method by comparing it to the  exact  solution that we can obtain when we
 restrict the Hilbert space dimension to such an extent that the exact diagonalisation is possible.

\section{Method}\label{metodo}
 In the Coupled-Cluster method\cite{reviews} one searches an approximated solution to the eigen-problem
   \begin{equation}
 H  \left| \Psi \right> =E  \left|  \Psi   \right>
\end{equation}
   where $H$ is the Hamiltonian in second quantization and is formed by a sum of products of one-particle 
creation and annihilation operators. The one-particle operators change, between $1$ and $0$,
 the occupation integer number of the one-particle orbitals contained in the model.

 The solution   is represented, given a reference state $\left| \Phi_0 \right>$, by the exponential ansatz
\begin{equation}
  \left| \Psi \right>= e^{S} \left| \Phi_0 \right> \simeq e^{S_N} \left| \Phi_0 \right>
\end{equation}
where $S$  is the ideal  exact solution and $S_N$ is a  sum, truncated to $N$ terms, of products of electron-hole pair excitations :
\begin{equation}
 S_N =  \sum^N_{i=1}{ t_i ~ Symm   { \left\{    \prod_{k=1}^{n_i} c^\dag_{\alpha_{i,k}}   c^\dag_{a_{i,k}}       \right\} }} \label{Ssum}
\end{equation}
 In this formula $N$ is the number of degrees of freedom of the ansatz. The larger is this number the more accurate is the representation. The $t_i$'s
are free coefficients that must  be obtained from the CCM equations below. Each term in the sum is the product of a set of electron(hole)-creation operators $c^\dag$. Each term
is determined by a choice of indexes $\alpha_{i,k}$( $a_{i,k}$), with the greek(latin) letter $\alpha$(a) ranging  
over empty(occupied) orbitals.
One can cosiders the reference state as the {\it vacuum} state and that each term $i$ in the sum $S_N$ creates, from vacuum, an excited state which is 
populated by $n_i$ particles (holes and/or electrons).
 The $Symm$ operator makes the ansatz symmetric for the Hamiltonian symmetry  subgroup 
which transforms, up to a factor, the reference state $\left| \Phi_0 \right>$ into itself. 

In the Coupled Cluster method, the rationale for the exponential ansatz resides in its size extensivity property. 
This means that for a system composed of two non correlated  parts, $A$ and $B$,
the  coupled cluster ansatz operator can be factorized as the product of two operators 
$e^{S^{A+B}}= e^{S^{A}}e^{S^{B}}$. This simple factorisability relation has deep consequences\cite{reviews}, 
whose one of the most important is that,  in a system with 
periodic translational symmetry,  the calculation complexity for a given accuracy
does not depends on the  system  size.

The CCM equations are obtained substituting $ \left| \Psi \right>$,   in the eigen-equation, with its ansatz
 and by multiplying at the left with $e^{-S_N}$, the inverse of the ansatz operator  .
One obtains for the eigenvalue
\begin{equation}
E =  \left< \Phi_0 \right|  e^{-S_N} H e^{S_N} \left| \Phi_0 \right> 
\end{equation}
while the free parameters are obtained setting the eigen-equation residue  to zero in the space
of  excited states which enter the $S_N$ sum :
\begin{equation}
0 =  \left< \Phi_0 \right|\left( \prod_{k=1}^{n_i}   c_{a_{i,k}}  c_{\alpha_{i,k}}  \right) e^{-S_N} H e^{S_N}  \left| \Phi_0 \right> ~ \forall i \in [1,N]  \label{ccmeqs}
\end{equation} 

The Coupled Cluster Method  expands these equations  by means of the  Hausdorff expansion formula
which  for two arbitrary operators $A$ and $B$ states that: 
\begin{equation}
  e^{-A}B e^{A}= B+[B,A]+1/2[[B,A],A]+..1/n![[B,A]...],A] +... \label{Hausdorff}
\end{equation}
The numerical applicability of CCM relies on the fact that, when $A$ is substituted with $S_N$ and $H$ replaces $B$, 
 only the first five terms in the series,  can be non zero. This can be demonstrated considering 
that  $S_N$ is formed by creation operators only,
 and that the interactions contained $H$  are composed 
by products of up to  four single-particle operators for the Coulomb interaction. For each term of the expansion
every $S_N$ entering in the commutators must have at least one one-particle creation operator contracted 
with one annihilation operator of $H$, for the term not to be identically zero.

Equation \ref{ccmeqs} gives $N$ polynomial equations by which we can determine the $N$ unknowns $t_i$.
These equations have order up to the fourth in the  $t_i$ variables because this is the maximum order in $S_N$
for the non-zero terms of the Hausdorff expansion. 

The number of solutions of a system of polynomial equations explodes exponentially with the 
number of equations and it is not possible, except for small systems, to explore systematically
the whole solutions space.
 To solve the equations, instead, we use  the Newton's method to follow the solution, increasing iteratively the number of free parameters and 
using as a starting point the for  $N$ parameters the $N-1$ ones found at the previous iteration plus a random choice 
for the $N^{th}$ one. 

The accuracy of the CCM solution increases with $N$. At each iteration 
the new $(N+1)^{th}$ term is constructed, in equation \ref{Ssum}, by assigning  its order $n_{(N+1)}$ and by choosing  
the concerned  electron and hole orbitals which are expressed by the sets of indexes $\alpha_{(N+1),k}$ and $a_{(N+1),k}$, with $k$ ranging from $1$ to $n_{(N+1)}$ .
We denote the ensemble of all possible choices with the symbol
\begin{equation}
\left\{    \left(  n^{\prime}_\zeta  ,\alpha^{\prime}_{\zeta,1}...,a^{\prime}_{\zeta,1}... \right)  \right\}
\end{equation}
where the possible choices satisfy the condition
\begin{equation}
0 \neq   \left< \Phi_0 \right|\left( \prod_{\zeta=1}^{n^{\prime}_\zeta}   c_{a^{\prime}_{\zeta,k}}  c_{\alpha^{\prime}_{\zeta,k}}  \right) e^{-S} H e^{S}  \left| \Phi_0 \right> \label{residue}
\end{equation}
The simplest choice consists in choosing the $\zeta$ which gives the largest residue in equation \ref{residue}. 

Once we have obtained the CCM ground state and its ground energy $E$, we are interested in the transition probability 
 for a time dependent perturbation
$exp(i \omega_D t) D$, where $D$ is an arbitrary product of $c^\dag$   operators.
The transition rate is given by the Fermi golden rule which states that the probability for the absorption of 
an energy quantum $\hbar \omega_D = \hbar \omega -E$, with $\hbar \omega$ being the final state energy, is proportional to:
\begin{equation}
  \rho_D(\omega,\gamma) =  \Im m   \frac{ \left< \Phi_0  \right|e^{S^\dag}   D^\dag  (H-\omega-i \gamma)^{-1} D e^{S}  \left| \Phi_0 \right>}
      { \left< \Phi_0  \right|e^{S^\dag}   e^{S}  \left| \Phi_0 \right>  }
\end{equation}
where $\gamma$ is a small line width. In order to calculate the above expression
we have to solve two problems : find an approximate solution $R$ for the  resolvent equation :
\begin{equation}
     (H-\omega-i \gamma) |R> =     D e^{S}  \left| \Phi_0 \right>    
\end{equation}
and calculate the scalar product.

We represent  an approximated  solution for the resolvent, introducing the approximating operator $R_{D,\omega,\gamma}$ and the following ansatz
 which is similar to the ansatz for $S$
with the difference that  it  contains both annihilation and creation 
operators   and that, in order to accede to the whole spectra, no symmetrization is done : : 
\begin{eqnarray}
  |R> =&  R_{D,\omega,\gamma} D e^{S}  \left| \Phi_0 \right>   \label{rsolution}\\
  R_{D,\omega,\gamma}  =& r^0_{D,\omega,\gamma} + \sum^{N^r}_{i=1}{ r^i_{D,\omega,\gamma}   \prod_{k=1}^{n^r_i} \hat c_{j_{i,k}}      } \label{ransatz}
\end{eqnarray}

in this expression $r^i$ are free parameters and we have introduced the  notation $\hat c$ to represent
in a compact way  both creation and annihilation operators.   The definition of the   $\hat c$
operator is, naming by  $N_{orbs}$ the total number of represented orbitals (occupied and empty) : 
\begin{equation}
  \hat c_j= \left\{ \begin{matrix}
    c^\dag_j   &  j \in [1,N_{orbs}]                          \\
    c_{j-N_{orbs}} &  j \in [N_{orbs}+1,2*N_{orbs}]                          
  \end{matrix} 
  \right|
\end{equation}

We build our spectral CCM equations (SCCM equations) by multiplying at the left with $e^{-S_N}$, and by setting the residue to zero :
\begin{eqnarray}
1 =&  \left< \Phi_0 \right|D^\dag  e^{-S} (H-\omega-i \gamma) R_{D,\omega,\gamma} D e^{S} \left| \Phi_0 \right> \\
0 =&  \left< \Phi_0 \right|D^\dag \left( \prod_{k=1}^{n^r_i} \hat c_{j_{i,k}} \right)^{*}  e^{-S}(H-\omega-i \gamma) R_{D,\omega,\gamma} D e^{S} \left| \Phi_0 \right> ~ \forall i \in [1,N^r]  \label{reseqs}
\end{eqnarray} 
Note that the validity of the above equations relies on the fact that $D$, being  a product of  $c^\dag$  operators, commutes with $S$.
These equations are expanded by the Hausdorff expansion formula substituting, in equation \ref{Hausdorff},  $A$ with $S_N$  and $B$ with $ (H-\omega-i \gamma) R_{D,\omega,\gamma} $. The  Hausdorff expansion contains, also in this case, a finite   number of non-zero terms because each term of the  the resolvent operator , in equation \ref{ransatz},
contains a finite number of annihilation operators while, as discussed above, each term of $H$ contains a maximum of four annihilation operators.

The expansion gives a set of linear equations  for the $r$ parameters. 
The resolvent equation accuracy is improved by systematically increasing $N^r$, selecting, at each iteration, the set of numbers
\begin{equation}
\left\{    \left(  n^{r}_{_{N^r+1}}  ,j_{_{N^r+1,0}},....,j_{_{N^r+1,n^r_{N^r+1}}} \right)  \right\}
\end{equation}
corresponding to the largest residue in the SCCM equations. When we calculate the residue we fix $\omega=\omega_r$
at the center of the spectral region of interest. Over the spectral region of interest the $r$ parameters are
given by a linear algebra operations of the kind  $ r=(M_1)^{-1} (M_2+\omega M_3) $ where the $M$'s are 
matrices obtained from SCCM expansion.
Once we know the $R$ operator we can calculate the spectra with the following equation :
\begin{equation}
  \rho_D(\omega,\gamma) =  \Im m   \frac{ \left< \Phi_0  \right|e^{S^\dag}   D^\dag   R_{D,\omega,\gamma} D e^{S}  \left| \Phi_0 \right>}
      { \left< \Phi_0  \right|e^{S^\dag}   e^{S}  \left| \Phi_0 \right>  } \label{Rspettro}
\end{equation}
This expression can be expanded using the Wick's theorem and the linked-cluster theorem as already done by Sourav et al.\cite{Sourav}. Contracting in all possible ways  the operators contained in  $  D ^\dag  R_{D,\omega,\gamma} D$  with themselves  one obtains sum of products of Green's function of different orders.
 To simplify this we use the simplest approximation which consists in setting to zero all the connected
Green's function excepted the one-particle Green's function :
\begin{equation}
      G(j_1,j_2) = \frac{ \left< \Phi_0  \right|e^{S^\dag} \hat c_{j_1} \hat c_{j_2}    e^{S}  \left| \Phi_0 \right>  }{ \left< \Phi_0  \right|e^{S^\dag}   e^{S}  \left| \Phi_0 \right>  } \label{greendef}
\end{equation}
We expand this equation for $G$  using   the Wick's and the linked-cluster theorems.
 We obtain a hierarchical set of equations involving Green's functions of arbitrary order. This expansion needs to be 
truncated choosing a closure relation. This closure relation is already provided by the choice that we have made   setting to zero all the connected Green's 
functions except  the two points one.
To obtain  the Dyson equation for the Green's function we proceed in the following way.
Each time we contract  a $\hat c$ operator with one of the terms contained in $S$, on the right, or with $S^\dag$ on the left,  a new vertex is obtained from which a number of new lines, equal to the term order minus one, are coming out. We consider all the combinatorial ways  of contracting all these  lines with themselves, except one branch 
which propagates further the Green's function.
This is the analogous of the Hartree-Fock approximation where two of the four legs of  each  Coulomb vertex 
are contracted with each other.

The  Dyson equation is solved iteratively.
The final result for the spectral function of equation \ref{Rspettro} depends linearly on the parameters $r$, functions of $\omega$,
and contain products of Green's functions ( defined by equation \ref{greendef}). 
The spectral resonances positions  depend on the $r$ parameters,  which 
are found by the SCCM equation and whose 
behavior  accounts for many-body correlations. The resonances  intensities, instead, 
depend on our Hartree-Fock-like truncation which still accounts many-body interactions
but in the mean-field approximation.

\section{Model}\label{modello}

In previous studies on manganites we applied exact diagonalisation, and Lanczos method, to the
study of resonant X-ray  scattering \cite{Mirone} at $L_2$,$L_3$ edges and $K_\beta$ fluorescence \cite{Herrero}.   
 The spectroscopy data were modeled with  a small planar cluster, described in  second quantization.
The model consisted of the  central Mn atom open shells orbitals, plus some selected orbitals
 localised on the first  neighbouring shell of oxygens atoms and Mn atoms.
These studies  revealed a  pronounced O 2p character of the doped charge carriers,  and 
the      non-local  nature of the  forces   governing  the charge redistribution phenomena which are very important in these systems.
The  accounting of few extra orbitals from  neighbouring shells, beside the resonating atom,  is crucial in describing these phenomena
 but one rapidly ecounters the  limit  of the  exponential growth of the Hilbert space dimension, when trying  to extend 
the size of the cluster. 
To calculate the ground state and spectra of larger systems, while still keeping a  good description
of the many-body correlations,  we have developed the methods described in this paper.

We will compare SCC method to exact numerical results that we will obtain in a truncated Hilbert space..
To keep the system numerically affordable for the exact diagonalisation technique  we consider a small  $2x2$ $MnO_2$ lattice with periodic boundary conditions. 
 The Mn sites are placed at integer coordinates $(2 i, 2 j)$ with $i$ and $j$ taking the values $0$ and $1$ , while
the oxygen atoms are at positions  $(2 i+1, 2 j)$  and  $(2 i, 2 j+1)$. 
In order to limit to the maximum the dimension of the Hilbert space we restrict the degrees of freedom
to those orbitals which are  the most important for the physics of manganites.
These are the $e_g$  $3d$ orbitals of $Mn$, namely  the $x^2-y^2$ and $3z^2-r^2$ orbitals,
 and the $p$ oxygen orbitals  which point toward $Mn$ sites.
 For oxygens
we restrict to $p_x$  for the  $(2 i+1, 2 j)$ sites and $p_y$ at  the $(2 i, 2 j+1)$ sites.
These are the oxygen orbitals which bridge the Mn sites along the $x$ and $y$ directions.
The system Hamiltonian is composed of several terms ;
\begin{equation}
	H = H_{bare}+H_{hop} +H^{Mn}_U +H^{Mn}_J +  H^{O}_U
\end{equation}
namely $H_{bare}$ which contains the one-particle energies of the orbitals, the hopping Hamiltonian $H_{hop}$
which moves electron between neighboring sites, the Hubbard correlations  $H^{Mn}_U$, 
$H^{Mn}_J$  and     $ H^{O}_U$  for Manganese and Oxygen.
The  $ H^{O}_U$ term  is used because,  applying   exact diagonalisation,  we truncate  the Hilbert space by limiting the  $p$ orbitals occupation numbers between
$1$ and $2$. In the CC method, instead, we cannot truncate because this  would destroy commutation relations.  
We have the possibility instead, in CCM, of choosing a high value of $U$ in the  Hubbard correlation  $ H^{O}_U$ , in conjunction with the oxygen part of  $H_{bare}$ to effectively limit the oxygen $p$ orbitals occupation numbers
thus making the comparison, with the truncated model exact solution,  possible.

The bare Hamiltonian is 
\begin{eqnarray}
H_{bare}=  \sum_{i,j,g_d,\sigma} \epsilon_{d,
\sigma}~  d^\dag_{_{g_d,\sigma,2i,2j}}d_{_{g_d,\sigma,2i,2j}} + \nonumber \\
(\epsilon_{p}-U_p)  \sum_{i,j,\sigma} \left( p^\dag_{_{x,\sigma,2i+1,2j}}p_{_{x,\sigma,2i+1,2j}} +p^\dag_{_{y,\sigma,2i,2j+1}}p_{_{y,\sigma,2i,2j+1}} \right)
\end{eqnarray}

where the $g_d$ index takes the values $ g_d=x^2-y^2,3z^2-r^2$, with $x,y$ being in plane and $z$ out of plane. The $Mn$ one-particle energies 
$\epsilon_{\sigma}$ are  spin-dependent to take into account  the mean-field exchange with 
the Mn $t_{2g}$ occupied orbitals  ($xy,xz,yz$)( whose degrees of freedom are discarded from the model).
The oxygen orbitals term takes into account the  Hubbard coefficient $-U_p$
to compensate  $ H^{O}_U$ and favoring double and single occupations on oxygens. 

The hopping term is 
\begin{equation}
 H_{hop} =  t~ \sum_{i,j,g_d,\sigma} \sum_{s=\pm 1}   s \left( f_{g_d,x}   p^\dag_{_{x,\sigma,2i-s,2j}} d_{_{g_d,\sigma,2i,2j}}  + 
 f_{g_d,y} p^\dag_{_{y,\sigma,2i,2j-s}} d_{_{g_d,\sigma,2i,2j}}  + c.c.
\right)
\end{equation}
where
\begin{eqnarray}
 f_{3z^2-r^2,x}=f_{3z^2-r^2,y}=1/2  \nonumber \\
 -f_{x^2-y^2,y}=f_{x^2-y^2,x}=\sqrt 3/2  \nonumber
\end{eqnarray}

The Coulomb intra-site repulsive interaction for $Mn$ is made by a part  for an electron pair on the same orbital,
and another part for two different orbitals : 
\begin{equation}
    H^{Mn}_U  \sum_{i,j,g_d} U_{d} ~  n_{_{g_d,\sigma=+\frac{1}{2},2i,2j}}n_{_{g_d,\sigma=-\frac{1}{2},2i,2j}} +
\sum_{i,j,\sigma_1,\sigma_2} U^\prime_{d} ~  n_{_{3 z^2-r^2,\sigma_1,2i,2j}}n_{_{x^2-y^2,\sigma_2,2i,2j}}
\end{equation}

The Coulomb exchange for $e_g$ orbitals is

\begin{equation}
    H^{Mn}_J =  J_{d}  ~  \sum_{i,j,\sigma_1, \sigma_2} d^\dag_{_{3 z^2-r^2,\sigma_2,2i,2j}}d^\dag_{_{x^2-y^2,\sigma_1,2i,2j}} d_{_{3 z^2-r^2,\sigma_1,2i,2j}}d_{_{x^2-y^2,\sigma_2,2i,2j}}
\end{equation}
 while the $e_g$-$t_{2g}$ exchange is included as a mean-field term inside $H_{bare}$. 

Finally the oxygen Hubbard term is 
\begin{equation}
      \sum_{i,j} U_{p} ~  \left( n_{_{p_x,\sigma=+\frac{1}{2},2i+1,2j}}n_{_{p_x,\sigma=-\frac{1}{2},2i+1,2j}} + n_{_{p_y,\sigma=+\frac{1}{2},2i,2j+1}}n_{_{p_y,\sigma=-\frac{1}{2},2i,2j+1}} +2 \right)
\end{equation}

The contributions of the terms factored by $U_p$, in the total Hamiltonian,  is identically zero when we restrict the  $n_p$ occupations  between  $1$ and $2$.

To fix the free parameters of the model we use knowledge from our previous work on manganites\cite{Mirone}. Parameters are given in $eV$ units. 
The effective Slater integrals used in that work correspond, in the present model, to   
$  U_{d}=6.88 $,  $ U^\prime_{d} = 5.049$, $J_{d}=-0.917$.  The exchange with occupied polarized $t_{2g}$ orbitals 
gives a $\simeq 2 eV$  splitting between $\epsilon_{d,\sigma=-\frac{1}{2}}=2$ and    $\epsilon_{d,\sigma=+\frac{1}{2}}=0$, in the case of ferromagnetic alignement.
We use a hopping $t=1.8$ taken from our previous work\cite{Mirone}.
The parameter $\epsilon_p$ controls  the amount of charge back-donation from oxygen to manganese. 
The predominant $O$ $2p$ character of doped holes found in manganites  \cite{Herrero} corresponds to a value $\epsilon_p$ which  
raises the bare oxygen orbitals energies above the bare  Mn ones. The value of $\epsilon_p$ influences the average occupation 
  of the $e_g$ orbitals.   These occupancies match the ones found in the previous works for a value $\epsilon_p \simeq 2$.

\section{Discussion}\label{discussione}

To find the CCM ground state and determine the resolvent equation we have adapted  our $Hilbert++$ \cite{hilbertxx,Mirone} code. This code  was originally created to calculate x-ray spectroscopies of small strongly correlated  clusters by exact diagonalisation. It implements a second quantisation representation of operators and determinants.
We have implemented  automatic computing of  commutators and  automatic extension of  the excitations set for CCM and for our SCC method. 

The exact diagonalisation and the Lancsoz tridiagonalisation for spectra calculation are performed with $Hilbert++$.
The code generates the  Hilbert space by  applying several times the Hamiltonian on a vector basis which is beforehand
 initialized   with a seed  state. In this seed state, the occupied spinorbitals are  all the oxygen ones and,
for ferromagnetic alignement on the $Mn$ sites, all the spinorbitals $3 z^2 -r^2$ with spin $\sigma=+1/2$ .
This state is named, in the rest of this paper,  {\it nominal reference configuration}.
The configurations having one or more oxygen sites unoccupied
, are discarded   in the exact calculation. With this limitation  on the  configurations, 
the generated Hilbert space growths up to a dimension which is slightly less than $7$ millions. 

To reproduce with the $CCM$ method the exact calculation done on the truncated space  we
set $U_p$  as high as $10^2eV$.

\begin{figure} 

\rotatebox{-0}{
\resizebox{1.0\columnwidth}{!}{
\includegraphics{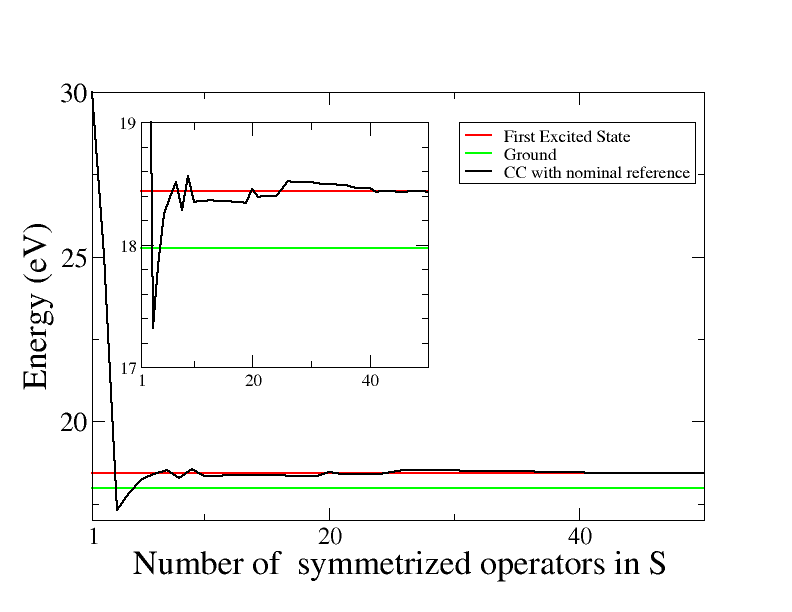}
}
}
\caption{\label{initialwf}  $CCM$ versus  exact diagonalisation of the  truncated Hilbert space space. 
For the  $CCM$ method we apply an effective truncation by setting  $U_p$  as high as $10^2eV$.
The convergence of the $CCM$ energy is shown as a function of the number of symmetrized eccitations contained in the $S$ operator.
The nominal reference generates the first excited state while the ground state has the symmetry of  $\left| \hat \Phi_0 \right>$,
equation \ref{groundsymm}
}
\end{figure}

We show in figure \ref{initialwf} the convergence of the $CCM$ energy as a function of the number of symmetrized eccitations contained in the $S$ operator
when we take the  {\it nominal reference configuration} as reference state 
The CCM energy converges, for the nominal reference, to the first excited eigenenergy, given by  exact diagonalisation, above the ground state.
We have analysed the ground  and the first excited states that we  obtain  by exact diagonalisation. The largest component of the first 
excited state is found to be  the nominal reference state. This explains why the CCM method, which takes this state as reference, converges to 
this eigen-state. The ground state, instead, has  a different symmetry. We find that there are four components which have the largest factor
and each of this component is obtained rotating   the $e_g$ electron,  on one of the four $Mn$ sites, from the $3 z^2 -r^2$ orbital to the $x^2-y^2$ one . More in details,
the ground state has the same symmetry of the state $\left| \hat \Phi_0 \right>$ given by 
\begin{equation}
\left| \hat \Phi_0 \right> = \sum_{i,j}     (-1)^{i+j}  d^\dag_{_{x^2-y^2,\sigma=1/2,2i,2j}} d_{_{3 z^2-r^2,\sigma=1/2,2i,2j}}  \left| \hat \Phi_0 \right> \label{groundsymm}
\end{equation}
 This state cannot be obtained starting from  the nominal reference with  the $CC$ method because it has completely different symmetry
properties. Notice for example that a $\pi/2$ rotation around the center of the cluster gives a factor $1$ if applied on
the nominal reference but  the same rotation gives a factor $-1$
when applied on   $\left| \hat \Phi_0 \right>$.

To force the $CC$ method to converge to such a symmetry state, a possible solution  would be using the multi-reference $CC$ method. We have not  implemented this method, which requires an a-priori knowledge of the solution,
 because our work  is focussed on our Spectral Cluster Method 
that, as we will see in this section, is able to detect these states exploring selected spectral regions.
To test further  the capabilities of CCM to converge to the true ground state
 we have allowed a possible  convergence to  $\left| \hat \Phi_0 \right>$ symmetry by using
a reference state $\left| \bar \Phi_0 \right>$ of lower symmetry : 
\begin{equation}
\left| \bar \Phi_0 \right> =   d^\dag_{_{x^2-y^2,\sigma=1/2,0,0}} d_{_{3 z^2-r^2,\sigma=1/2,0,0}}  \left|  \Phi_0 \right>
\end{equation}

\begin{figure} 
\rotatebox{-0}{
\resizebox{1.0\columnwidth}{!}{
\includegraphics{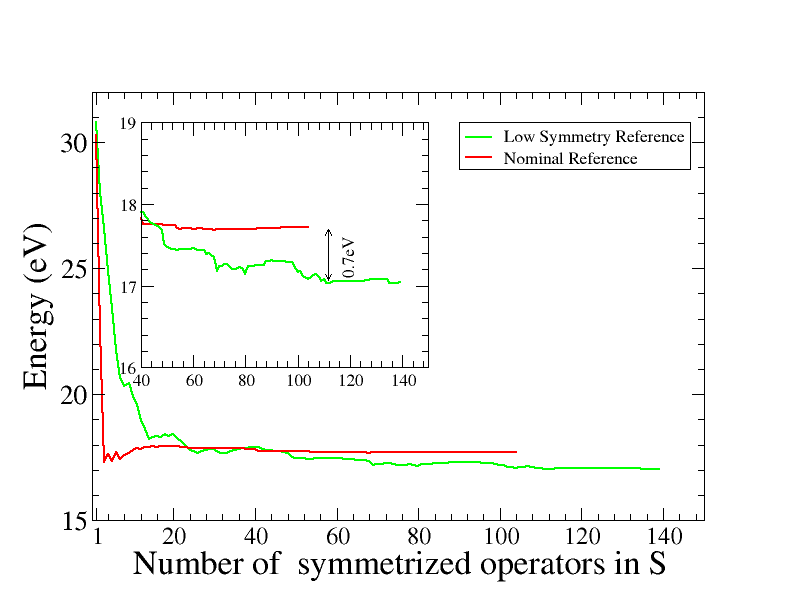}
}
}

\caption{\label{initialwf_lowU} Convergence of the  $CCM$ energies for a non-truncated Hilbert space ($U_p \simeq 5eV$). 
The nominal reference generates the first excited state while the  lower symmetry reference state  $\left| \bar \Phi_0 \right>$
gives the ground state.
}
\end{figure}

We show in figure  \ref{initialwf_lowU}  the convergence of the CCM energies for $U_p=5eV$  using two different choices of the reference state : the high symmetry state  $\left|  \Phi_0 \right>$  and the lower symmetry  $\left| \bar \Phi_0 \right>$ .
The CCM solution for reference state $\left| \bar \Phi_0 \right>$ converges to a lower energy than the one obtained with the nominal reference state.
We cannot compare this calculation, done with $U_p=5eV$, with the results of exact diagonalisation because the small value of $U_p$
gives access to a larger Hilbert space which is computationally more expensive.
On the other hand  for an high value of $U_p=10^2eV$, when comparaison with exact diagonalisation is possible, we have not been able to  
obtain the ground state starting from the low symmetry reference state $\left| \bar \Phi_0 \right>$.
We think that this difficulty can be explained in the following way : the lower energy of the $\left| \hat \Phi_0 \right>$ symmetry state
is due, in the $CCM$ equations, to a kind of  bridges, made of operators which link  the components  of  $\left| \hat \Phi_0 \right>$ to each other.
These bridges are created when an excitation operator, which composes $S$,   is transformed by the Hausdorff  commutation expansion 
into another excitation operator which will subsequently enter $S$, and so on. For the particular symmetry of    $\left| \hat \Phi_0 \right>$
to be obtained from    $\left| \bar  \Phi_0 \right>$,   these bridges must be long enough to transform one component into another.
The problem of using a high value of $U_p$ is that for every pair of excitation operators which both create  a hole 
on the same oxygen  site, a new term coming from their product, will appear in the residue containing  two holes that site. This will necessitate a new higher order excitation to be   subsequently  included in $S$  whose contributions will cancel the product
of the two operators. This because the very high value of $U_p$ forbids double hole occupancies on oxygen sites.
The need of accounting more operators requires more iterations. During these iteration the  $\left| \hat \Phi_0 \right>$ symmetry is non favorable and our procedure converges to higher eigenvalues. The CCM wavefunction corresponding  to the true ground state becomes
energetically favorable for a number  of eccitation operators about $40$. 
 The use of a  multireference could have been used to force a particular symmetry. This procedure  would have been feasible 
for the small system that we have treated in this work, because we can know the ground state symmetry
from the exact solution. For larger systems, however, even if one could a-priori know  the correct symmetry,
the number of determinants in the multireference state  grows exponentially with the size of the system.
Moreover, the Newton method, used for solving the CCM equations, does not guarantee that the
lowest energy solution will be found, because this method allows to follow just one solution which might  be not the good one.

The spectral method that we present in this work allows instead to explore, focussing on
selected spectral regions, a larger set of solutions which are observed as resonances.
  
We show in figure 3 the spectra for the first excited eigenvalue at $\epsilon_p=2eV$, $U_p=10^2eV$, considering a probe operator  $D_{cf}$
which induces a crystal-field rotation in the $e_g$ space, on one $Mn$ site :
\begin{equation}
  D_{cf}= d^\dag_{_{x^2-y^2,\sigma=1/2,0,0}} d_{_{3 z^2-r^2,\sigma=1/2,0,0}}
\end{equation}
The SSCM equations reproduces well the exact diagonalisation results.
The spectra shows a peak at negative energy. This is the  ground state which was not accessible 
starting from the nominal reference state but it is visible as a resonance in the SCCM equations.
The SCCM residues, to expand the $R$ operator, have been calculated fixing $\omega_r$ at zero
because crystal field excitations are found at low energies. The SCCM spectra has been calculated with 
$N^r=10^4$.

Figure 4 shows the spectra for the same initial state, but for a probe operator which 
transfers charge from a oxygen site to a neighbouring  manganese: 
\begin{equation}
  D_{cf}= d^\dag_{_{x^2-y^2,\sigma=+1/2,0,0}} p_{_{p_x,\sigma=+\frac{1}{2},1,0}} 
\end{equation}
The SCCM spectra has been calculated considering two energy windows :
  one around the charge transfer peak using   $\omega_r=3eV$,$N^r=8  \times 10^3 $ , and another
window around the ground state, using $\omega_r=-0.5eV$,$N^r= 10^4 $.

Figure 5 show the same spectra calculated for a non-truncated Hilbert space, using $U_p=5eV$.
Comparison to exact calculation is not possible in this case but we can see that the most  important features are preserved, namely the charge transfer peak, the ground state peak and the peak due to the overlap with the initial state at zero absorbed energy.
The convergence of the spectra in this case of low $U_p$ is easier and the spectra can
be calculated with only one energy window, using  $\omega_r=3eV$.
The graph shows two curves, one calculated with $N^r= 10^3$, where the charge transfer peak
is already in place, and another done at a higher value of $N^r= 7 \times 10^3$
which is necessary to have a proper convergence on the ground state peak at $\simeq -0.7eV$. 
The different behaviour for the two peaks can be seen as a consequence of the 
non-locality of the ground state derived from $\left| \hat \Phi_0 \right>$  symmetry. The non-locality  implies
a larger set of terms entering the resolvent sum. 

\section{Conclusions}

We have applied CCM equations to a strongly correlated lattice in the case of strong departure from the reference state.
We have developped the spectral coupled cluster equations,  by 
finding an approximation to the resolvent operator,  that gives the spectral response for the class of  probes
that are writable as products of creation operators

We have applied the method to a $MnO_2$ plane model for a parameters choice which makes the  ground state
particularly difficult to find with the CCM equations because of its peculiar symmetry which corresponds to a non-nominal reference state. 
 We have shown that this state can be   spectrally observed using SCCM equations
by probing a CCM solution for the nominal reference state. In this case one observes a negative 
energy solution which corresponds to the true ground state.
We think that  CCM and SCCM equations have a strong potential, for strongly correlated lattices, not only for the study
of the ground state but also for all those excitations that can be represented by a resolvent operator $R$ that can be written as the sum of localised terms.  

\section{Acknoweledgment}

I dedicate this work to the memory of my father Paolo.
 I acknoweledge Javier Fernandez Rodrigues who helped me in  setting up  the exact diagonalization of the $MnO_2$ plane  model during a post-doctoral 
stage financed by the {\it Gobierno del Principado de Asturias}  in the frame of the 
{\it  Plan de Ciencia, Tecnologia e Innovacion PCTI de Asturias 2006-2009.}.
I thank Markus Holzmann for critically reading the paper.

\newpage

\begin{figure} 
\rotatebox{-0}{
\resizebox{1.0\columnwidth}{!}{
\includegraphics{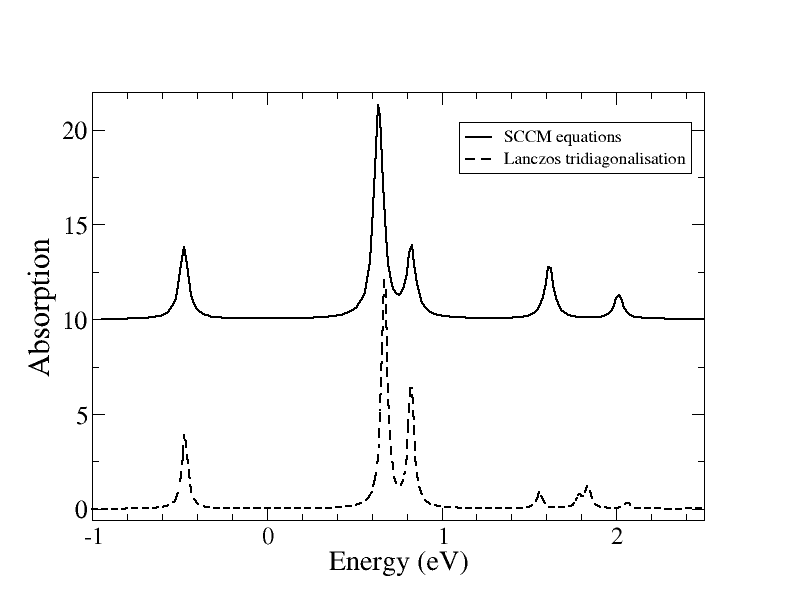}
}
}

\caption{\label{spettrocf}
  Spectra for the out-of-plane to in-plane  probe  $D_{cf}$ applied on  the first excited eigenvalue, $U_p=10^2eV$ (truncated hilbert space)
}
\end{figure}

\begin{figure} 
\rotatebox{-0}{
\resizebox{1.0\columnwidth}{!}{
\includegraphics{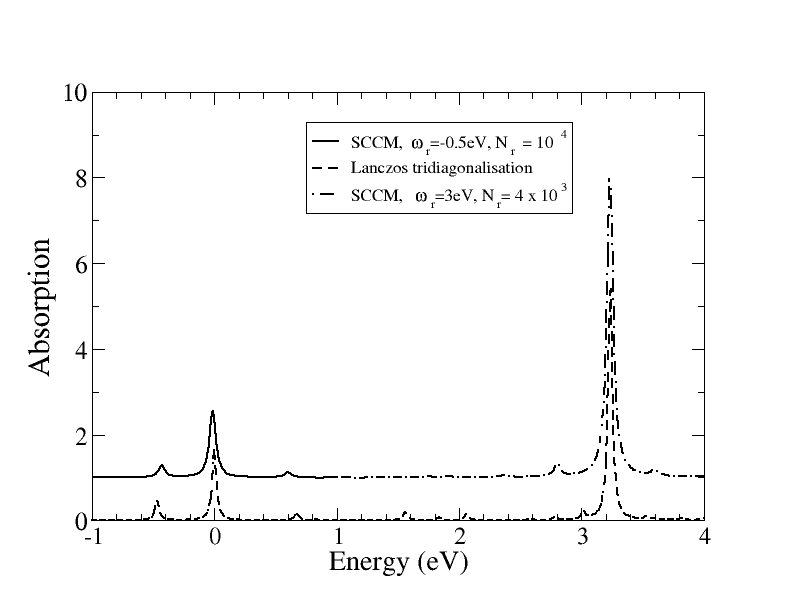}
}
}

\caption{\label{spettroch} 
  Spectra for the o charge transfer  probe  $D_{ch}$
 applied on  the first excited eigenvalue of the truncated problem ( $U_p=10^2eV$ )
}
\end{figure}

\begin{figure} 
\rotatebox{-0}{
\resizebox{1.0\columnwidth}{!}{
\includegraphics{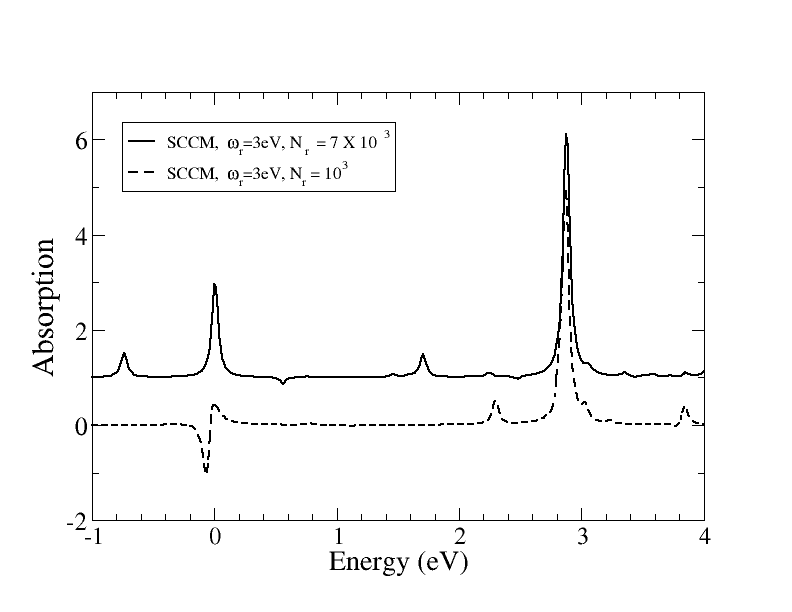}
}
}

\caption{\label{spettrochlow} 
  Spectra for the charge transfer  probe  $D_{ch}$
  applied on  the first excited eigenvalue of 
the non truncated problem ( $U_p=5eV$ ).
}
\end{figure}

\end{document}